\documentclass[12pt,preprint]{aastex}

\usepackage{graphicx}
\usepackage{t1enc}
\usepackage[varg]{txfonts}
\usepackage{epsfig}
\usepackage{rotating}
\usepackage{natbib}
\usepackage{color}

\newcommand{\kms}   {km~s$^{-1}$}

\newcommand{\jpb}   {$\rm Jy~beam^{-1}$}    
\newcommand{\lo}    {$L_{\sun}$}

\newcommand{\et}    {et al.}
\newcommand{\eg}    {e.\,g.,}

\newcommand{\hii}   {H{\small II}}
\newcommand{\uchii} {UC~H{\small II}}
\newcommand{\hchii} {HC~H{\small II}}
\newcommand{\raun}  {$^\mathrm{h~m~s}$}
\newcommand{\deun}  {$\mathrm{\degr~\arcmin~\arcsec}$}
\newcommand{\supa}  {$^\mathrm{a}$}
\newcommand{\supb}  {$^\mathrm{b}$}
\newcommand{\supc}  {$^\mathrm{c}$}
\newcommand{\supd}  {$^\mathrm{d}$}
\newcommand{\supe}  {$^\mathrm{e}$}
\newcommand{\supf}  {$^\mathrm{f}$}
\newcommand{\supg}  {$^\mathrm{g}$}
\newcommand{\phnn}  {\phantom{0}\phantom{0}}

\definecolor{RED}{rgb}{1.0,0.0,0.0}

\shorttitle{Searching for new hypercompact \hii\ regions}
\shortauthors{S\'anchez-Monge et al.}

\begin{document}

\title{Searching for new hypercompact \hii\ regions}

\author{\'Alvaro S\'anchez-Monge\altaffilmark{1,2},
	Jagadheep D. Pandian\altaffilmark{3},
	and
	Stan Kurtz\altaffilmark{4}}
\email{asanchez@arcetri.astro.it}
\altaffiltext{1}{Osservatorio Astrofisico di Arcetri, INAF, Largo E.\ Fermi 5,
I-50125 Firenze, Italy}
\altaffiltext{2}{Departament d'Astronomia i Meteorologia (IEEC-UB), Institut de
Ci\`encies del Cosmos (ICC), Universitat de Barcelona, Mart\'i i Franqu\`es, 1,
E-08028 Barcelona, Spain}
\altaffiltext{3}{Institute for Astronomy, University of Hawaii, 2680 Woodlawn
Dr., Honolulu, HI 96822, USA}
\altaffiltext{4}{Centro de Radioastronom\'ia y Astrof\'isica, Universidad
Nacional Aut\'onoma de M\'exico, Apdo.\ Postal 3-72, 58090, Morelia, Michoac\'an,
Mexico}

\begin{abstract}
Hypercompact (HC) \hii\ regions are, by nature, very young \hii\ regions,
associated with the earliest stages of massive star formation. They may
represent the transition phase as an early B-type star grows into an O-type
star. Unfortunately, so few \hchii\ regions are presently known that their
general attributes and defining characteristics are based on small number
statistics. A larger sample is needed for detailed studies and good statistics.
Class II methanol masers are one of the best indicators of the early stages of
massive star formation. Using the Arecibo Methanol Maser Galactic Plane Survey
--- the most sensitive blind survey for 6.7~GHz methanol masers to date --- we
selected 24  \hchii\ region candidates. We made EVLA continuum observations at
3.6 and 1.3 cm to search for \hchii\ regions associated with these masers. We
identified six potential \hchii\ regions in our sample based on the presence of optically
thick free-free emission. Overall, we find that 30\% of the methanol masers
have an associated centimeter radio continuum source (separation less than 0.1~pc), 
which is in general agreement with previous studies.
\end{abstract}


\section{Introduction}

Massive stars are key to understanding many physical phenomena in the Galaxy;
nevertheless, their formation process is still poorly understood. A prime
manifestation of recent massive star formation is the presence of an \hii\
region surrounding the young massive star(s). The range of sizes and densities
of these \hii\ regions reflects both the original conditions of the star-forming
environment and the later evolution of the ionized regions. The smallest
($<\!0.03$~pc) and densest ($n_\mathrm{e}>\!10^{6}$~cm$^{-3}$) \hii\ regions,
the so-called hypercompact (HC) \hii\ regions (\eg\ Kurtz 2005), are associated
with the earliest stages of evolution. \hchii\ regions are particularly
interesting in the study of massive star formation because they are so small
that we can assume they harbor only a \emph{single star} (or a binary system),
in contrast to ultracompact (UC) or compact \hii\
regions (with sizes $\sim\!0.5 -- 1$~pc) which usually harbor clusters of massive stars
(\eg\ Kurtz 2000; Hoare \et\ 2006). Additionally, recent work (Keto 2007)
suggests that O stars form gradually via accretion through an \hchii\ region
onto intermediate mass B stars. Thus, the study of \hchii\ regions is important
to understand the formation of massive stars. However, only a handful of \hchii\
regions are known, and a larger sample is needed for detailed studies with good
statistics to determine the properties of this very early stage of massive star
formation.

The $5_{1}$--$6_{0}$ $A^{+}$ line of methanol at 6.7~GHz is the strongest of the
class~II methanol masers. Theoretical and observational studies strongly suggest
that this maser line traces very early stages of massive star formation ---
often prior to the formation of an \uchii\ region (\eg\, Minier \et\ 2001; Cragg
\et\ 2005; Ellingsen 2006; Schnee \& Carpenter 2009). Unlike water, OH or
class~I methanol masers, that have also been found toward low-mass star forming
regions, class~II methanol masers seem to be unique to {\it massive} young
stellar objects (YSOs). Thus, class~II methanol masers are excellent candidates
to search for \hchii\ regions.

The Arecibo Methanol Maser Galactic Plane Survey (AMGPS; Pandian \et\ 2007) is
the most sensitive blind survey for 6.7~GHz methanol masers carried out to date.
By using the 305~m Arecibo radio telescope, the AMGPS covered an area of
18.2~deg$^2$ between Galactic longitudes of 35\degr\ and 54\degr, detecting a
total of 86 sources. Thus, this catalog is an excellent starting point to look
for new \hchii\ region candidates.

\section{Selection of Targets}

\hchii\ regions have small sizes ($<0.03$~pc), high densities
($>10^{6}$~cm$^{-3}$), high emission measures ($>10^{10}$~pc~cm$^{-6}$), steep
spectral indices ($\sim\!+1$), and broad ($>40$~\kms) radio recombination lines
(\eg\ Gaume \et\ 1995; Kurtz 2000; Sewilo \et\ 2004). The turnover frequency,
separating the optically thin ($S_{\nu}\propto\nu^{-0.1}$) and thick
($S_{\nu}\propto\nu^{+2}$) regimes is usually between 20 and 40~GHz for \hchii\
regions (substantially higher than the typical 5~GHz for \uchii\ regions).
Hence, \hchii\ regions are faint at wavelengths longer than  6~cm, but their
intensities increase at shorter wavelengths. In order to select those AMGPS
sources with the highest probability of harboring an \hchii\ region, we
developed the following selection criteria: \emph{i)} \emph{no 6~cm~/~20~cm
counterparts}, we searched the NRAO VLA Sky Survey (NVSS, Condon \et\ 1998), and
the MAGPIS and CORNISH surveys (Helfand \et\ 2006; Purcell \et\ 2008) for
emission at 20 and 6~cm toward all the AMGPS methanol maser sources. By
rejecting sources with 20 or 6~cm emission we eliminate star-forming regions
with more-developed, lower-density \hii\ regions, detectable at longer
wavelengths; and \emph{ii)} \emph{mid-IR emission}, we searched the MSX (Price
et al. 2001) and MIPSGAL  (Carey et al. 2009) catalogs for emission at
mid-infrared wavelengths toward those sources with no 20~cm/6~cm emission.
Essentially all sources show emission in the MIPSGAL 24~$\mu$m band (in
some cases with pixels blanked due to saturation), but only half of the sources
were detected by MSX. This is primarly because of the significantly better
sensitivity of {\it Spitzer} compared to MSX. However, MSX detections suggest
the presence of a relatively warm young stellar object, which may be more likely
to host a young \hii\ region. We hence rejected sources that have no MSX 
detection. Thus all our targets exceed the MSX detection limit of
$\sim$0.3~Jy at 8~$\mu$m.

Of the 86 methanol masers in the AMGPS catalog, 24 fulfill these selection
criteria. The peak flux densities of the selected masers range from 0.27~Jy to
17~Jy, and their luminosities (taken from Pandian \et\ 2009) range from
$9.0\times10^{-9}$ to $2.6\times10^{-5}$~\lo, covering the full range of flux
densities and luminosities reported in the AMGPS catalog. Thus, we can test for
correlations between the maser properties and the presence of \hchii\ regions.

\section{Observations and Data Reduction}

Our sample was observed with the EVLA (see Perley \et\ 2011) of the
NRAO\footnote{The Expanded Very Large Array (EVLA) is operated by the National
Radio Astronomy Observatory (NRAO), a facility of the National Science
Foundation operated under cooperative agreement by Associated Universities,
Inc.} in the continuum at 3.6~cm and 1.3~cm. The observations were made during
the shared-risk phase in 2010 April and June; the array was in the D
configuration. Owing to missing scans during the shared-risk observing not all
sources were observed at both frequencies: 19 regions were observed at 3.6~cm,
and the entire sample (24 regions) at 1.3~cm. The new Wide-band Interferometric
Digital Architecture (WIDAR) correlator provided continuum observations of
128~MHz bandwidth with two polarizations, resulting in an effective bandwidth of
256~MHz. On-source integration times were about 6~minutes at 3.6~cm and
10~minutes at 1.3~cm.  Flux and bandpass calibration were performed by observing
the quasar 3C48, with an adopted flux density of 3.15~Jy at 3.6~cm and 1.12~Jy
at 1.3~cm. Amplitude and phase calibrations were achieved by monitoring the
quasars J1851$+$0035 and J1922$+$1530, resulting in an rms phase of 5\degr\ and
50\degr\ at 3.6~cm and 1.3~cm, respectively. Data were calibrated using the CASA
package, following the standard guidelines for continuum reduction. The
calibrated \emph{uv}-data sets were exported in FITS format, and imaged using
the NRAO package AIPS. Images were made using a variety of weightings, and
tapering the 1.3~cm \emph{uv}-data to improve the sensitivity to
low-surface-brightness objects.  Typical synthesized beams were
$\sim\!9$\arcsec\ and $\sim\!6$\arcsec\ at 3.6~cm and 1.3~cm, respectively, with
rms noise levels of $\sim\!30$~$\mu$\jpb\ and $\sim\!100$~$\mu$\jpb.

\section{Results: centimeter continuum sources}

Radio continuum emission is seen in all the regions. At 3.6~cm we detect
continuum sources in all 19 observed fields, while at 1.3~cm we detect continuum
emission in 10 out of 24 fields. A total of 49 radio continuum sources
were identified in the 24 fields of our sample (2 out of 49 sources were
detected at 1.3~cm in fields that were not observed at 3.6~cm). However, not all
the sources are found close to the methanol maser emission. A complete analysis
of all the sources (including those located far from the maser emission) with a
multi-wavelength (infrared, submillimeter and centimeter) analysis will be
presented in a forthcoming paper.

In this Letter, we focus our attention on the centimeter continuum sources close
to the methanol maser emission. In particular, we study the sources detected
within the 1.3~cm primary beam ($\sim\!2$\arcmin), which is centered (in most
cases) at the coordinates of the methanol masers reported by Pandian \et\
(2011). Within the 1.3~cm primary beam, 11 of the 24 fields show radio continuum
emission.  Thus, the detection rate of centimeter continuum  sources close to
the maser emission ($<1$\arcmin) is around 46\%. A total of 14
sources were detected in these 11 fields.

Table~\ref{t:cmmaser} lists the 24 regions observed in this study. For regions
with a source within the 1.3~cm primary beam we list the position of peak
emission (derived from the 3.6~cm image), and the primary beam corrected flux
densities at 3.6 and 1.3~cm. An upper limit of 4 times the rms noise of the map
is listed if there is no detection within the 1.3~cm primary beam. We also list
the properties  of the 6.7~GHz methanol masers (peak flux density and
luminosity) from  Pandian \et\ (2007, 2009), as well as the angular and spatial
offset between the  centimeter continuum sources and the maser spots.

The centimeter continuum emission found close to the masers is typically compact
with respect to our synthesized beams of 6\arcsec--9\arcsec, except for two
sources in G38.26$-$0.20 and G48.99$-$0.30, for which we found extended
structures that are offset from the maser positions and are partially resolved
out at 1.3~cm. The maximum angular scale that can be recovered by our
observations is $\sim\!170$\arcsec\ and $\sim\!65$\arcsec\ at 3.6 and 1.3~cm,
respectively. However, the snapshot nature of the observations make it difficult
to properly image large structures. To obtain more accurate estimates of the
spectral indices for the compact sources (see Table~\ref{t:cmmaser}) we
re-imaged these fields at 3.6 and 1.3~cm using the \emph{uv}--range
3--28~k$\lambda$. Flux densities from this restricted \emph{uv}--range were
used to calculate the spectral indices reported in Table~\ref{t:cmmaser}.

\begin{figure*}[t]
\begin{center}
\begin{tabular}[b]{c c c}
\noalign{\smallskip}
	\epsfig{file=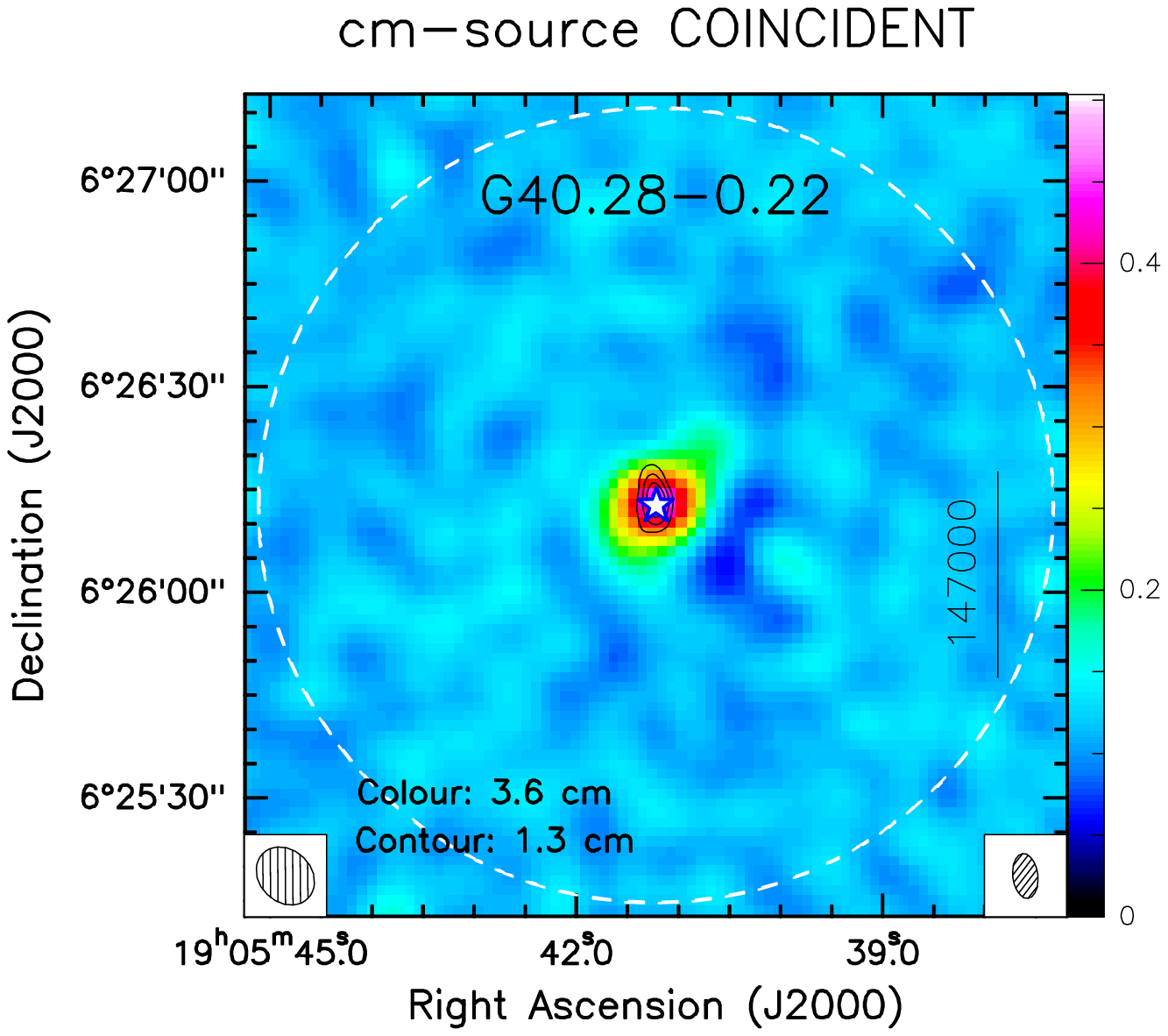, scale=0.6}	&&
	\epsfig{file=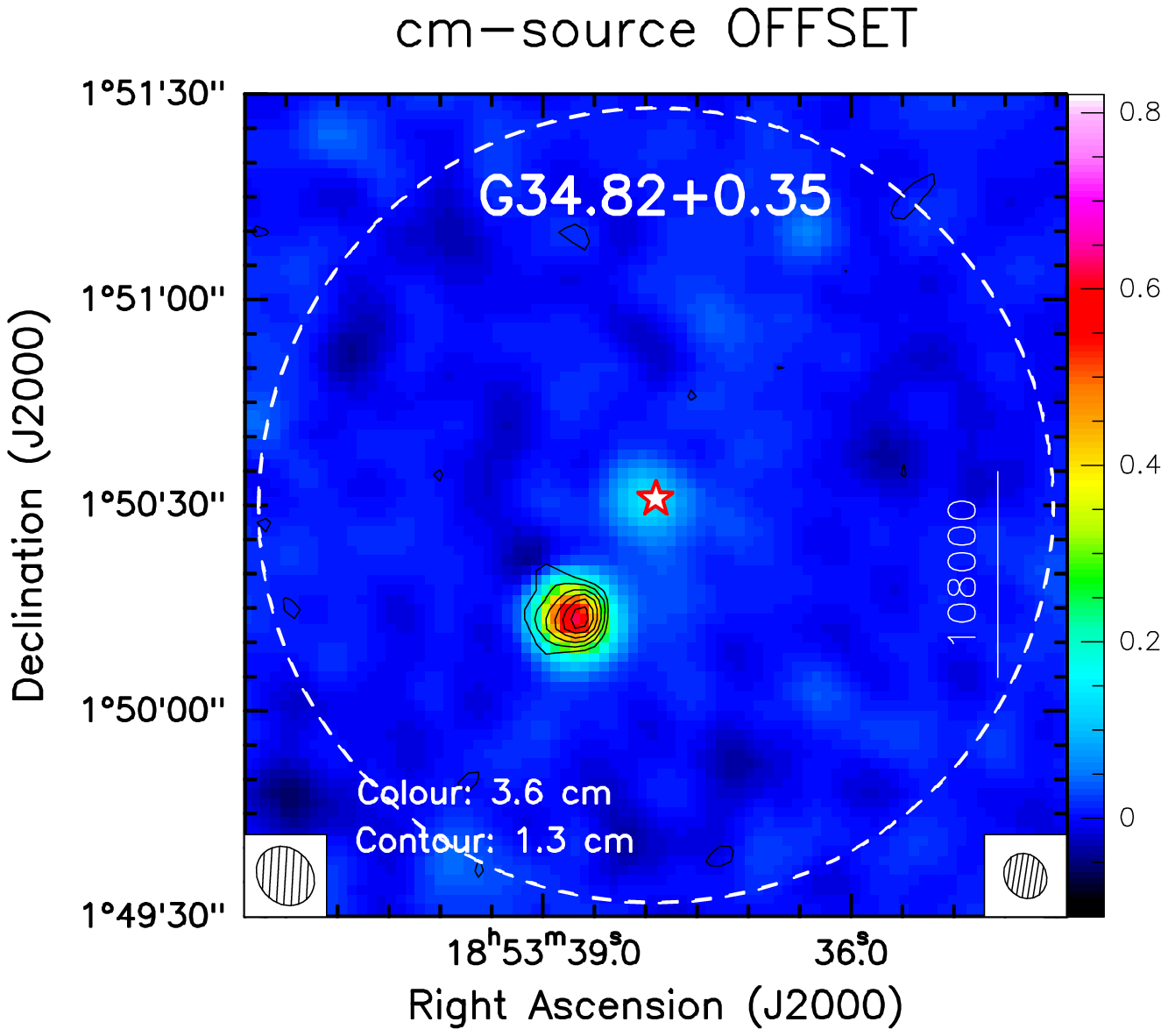, scale=0.6}	\\
\noalign{\smallskip}
\noalign{\smallskip}
\noalign{\smallskip}
\noalign{\smallskip}
	\epsfig{file=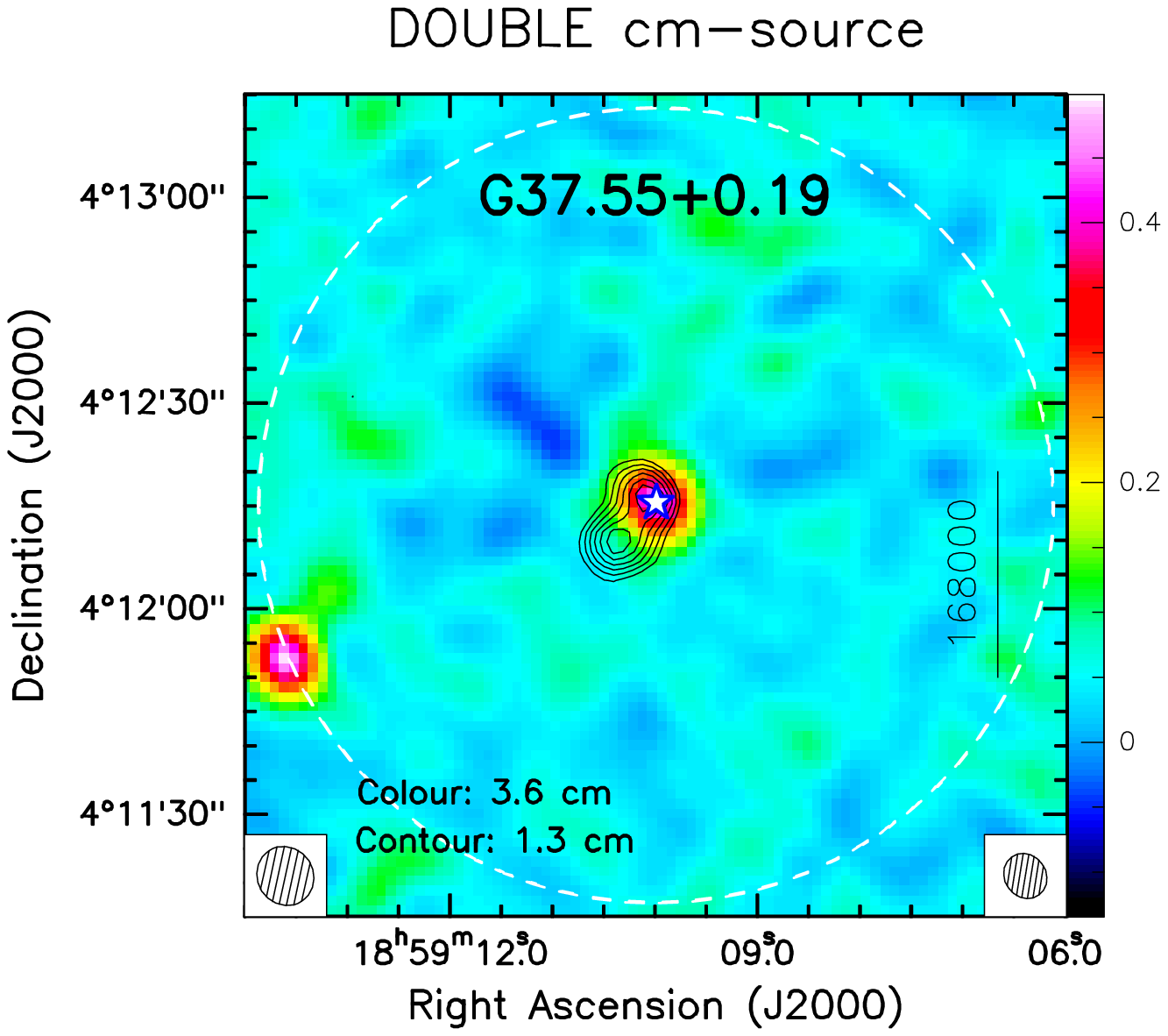, scale=0.6}	&&
	\epsfig{file=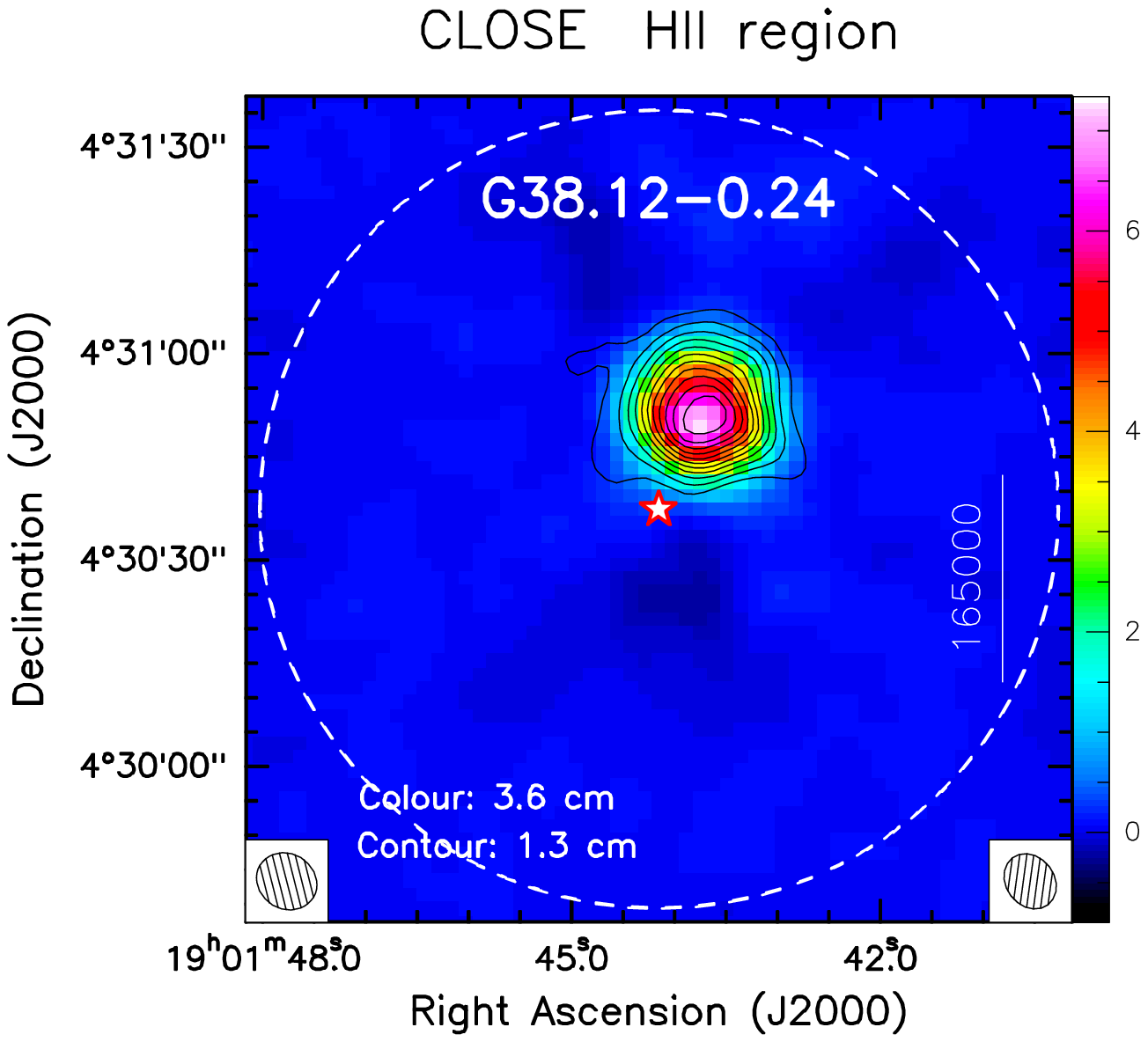, scale=0.6}	\\
\end{tabular}
\caption{Continuum images. Color scale: 3.6~cm VLA continuum image. The scale is
indicated in the (right) color bar of each plot (in mJy~beam$^{-1}$). Black
contours: 1.3~cm VLA continuum image. G40.28$-$0.22: levels are 3 to 15 in steps
of 2, times the map rms of 74~$\mu$\jpb. G34.82$+$0.35: levels are 2 to 7 in
steps of 1, times the map rms of 88~$\mu$\jpb. G37.55$+$0.19: levels are 3
to 8 in steps of 1, times the map rms of 75~$\mu$\jpb. G38.12$-$0.24: levels
are 3 to 33 in steps of 3, times the map rms of 120~$\mu$\jpb. Synthesized
beams are shown in the bottom-left (3.6~cm) and bottom-right (1.3~cm) corners.
The red star marks the position of the methanol maser (Pandian \et\ 2011). The
primary beam at 1.3~cm is indicated with the white dashed circle. The spatial
scale is indicated in astronomical units.}
\end{center}
\label{f:cmmaps}
\end{figure*}

Most of the compact sources show positive spectral indices ($> +0.3$), 
indicative of partially optically thick emission typical of \hchii\ regions and
thermal radiojets (\eg\ Kurtz 2005; Rodr\'iguez \et\ 2005). A few sources have
spectral indices between $-0.3$ and $-0.2$, possibly tracing optically thin
emission from weak and compact \hii\ regions (\eg\ S\'anchez-Monge \et\ 2008),
or non-thermal emission as found toward other massive young stellar objects
(\eg\ Zapata \et\ 2006). Two sources in our sample were detected only at
1.3~cm: the south-eastern source in G37.55+0.19 (see Figure~\ref{f:cmmaps}) and
the source in G53.14+0.07. Adopting a 4$\sigma$ detection limit at 3.6~cm, we
can place a lower limit on the spectral index of these sources. The former has
$\alpha>1.3$ while the latter has $\alpha>1.7$. These large values for the
spectral index indicate that the turnover frequency (from optically thick to
optically thin emission) is greater than 22~GHz, which implies a lower limit to
the emission measure of $2\times10^9$~pc~cm$^{-6}$. These two sources are
excellent candidates for \hchii\ regions. Several of the regions we observed
were previously studied with the VLA at the higher angular  resolution of
$\sim\!1$\arcsec~(Pandian \et\ 2010). Our results  are similar to those of 
Pandian \et\ (2010), although we typically recover more flux in our
D-configuration observations.

To describe the morphology and spatial distribution of the emission with respect
to the masers, we classify the sources into three main groups: \emph{a)}
compact sources spatially associated with the methanol maser (7/14, 50\%),
\emph{b)} compact sources offset  from the maser (2/14, 14\%), and
\emph{c)} larger (in some cases, clearly extended) sources found offset from the
maser emission, likely tracing more evolved UC \hii\ regions (5/14, 36\%).
In Figure~\ref{f:cmmaps} we show an example of each group.

\section{Discussion}

Of the 14 sources detected within the 1.3~cm primary beam (see Section~3
and Table~\ref{t:cmmaser}), seven are found to be coincident (spatial offset
$<0.1$~pc) with a methanol maser. Considering the entire sample of 24 distinct
maser sources, 29\% of the methanol masers are associated with radio continuum
emission with a flux density $\gtrsim\!0.4$~mJy. This is consistent with
previous work: 10\% of the AMGPS masers has 20~cm NVSS counterparts (Pandian
\et\ 2007b), 33\% of the Torun blind survey (Szymczak \et\ 2002) shows
associated 6~cm emission, and 30\% of the AMGPS sub-sample studied by Pandian
\et\ (2010) was found to have associated centimeter continuum emission.

It is interesting to note that the sources associated with maser emission are
compact and all but one have spectral indices between $+0.3$ and $+1.7$ with an
average value of $+0.6$, indicating the presence of partially optically thick
emission. In contrast, centimeter continuum sources found offset ($>0.1$~pc) 
from the maser position are generally morphologically more extended and have
optically thin emission (see Table~\ref{t:cmmaser}). This is consistent with
6.7~GHz methanol masers tracing very early phases of massive star formation. We
do not see any clear correlation between the properties of the centimeter
continuum emission and those of the maser emission (peak flux density or
luminosity). However, we caution that our sample is small; a more extensive
survey might show some correlation between the continuum and maser emission.

We determined the physical parameters of the centimeter continuum sources
assuming they are (homogeneous) \hii\ regions. For those sources with optically
thin emission (G38.12$-$0.24, G38.92$-$0.36, and G42.70$-$0.15) we derive an
emission measure of $\sim\!10^4$--$10^5$~cm$^{-6}$~pc, characteristic of \uchii\
regions (\eg\ Wood \& Churchwell 1989), and an ionizing photon flux of
$\sim\!10^{46}$~s$^{-1}$ (indicating ionization from B0 or B0.5 stars, Panagia
1973). For sources with optically thick emission, we performed a similar
analysis to that done by Pandian \et\ (2010). Using the Altenhoff \et\ (1960)
approximation, the free-free optical depth can be related to the emission
measure, $EM$, as
\begin{equation}
\tau_\nu=
0.08235~\bigg[\frac{EM}{\mathrm{pc~cm}^{-6}}\bigg]~
\bigg[\frac{T_\mathrm{e}}{\mathrm{K}}\bigg]^{-1.35}~
\bigg[\frac{\nu}{\mathrm{GHz}}\bigg]^{-2.1};
\end{equation}
here we assume an electron temperature of $10^4$~K. Given that the spectral
index is $\gtrsim\!+0.3$, and considering homogeneous \hii\ regions, the
emission measure should be $\gtrsim\!10^9$~cm$^{-6}$~pc (with the limiting case
when $\tau_\nu=1$ at a frequency of 22.46~GHz). In the optically thick regime,
we can relate the flux density and the solid angle of the source,
$\Omega_\mathrm{S}$, as
\begin{equation}
S_\nu=
B_\nu(T_\mathrm{e})~\Omega_\mathrm{S},
\end{equation}
with $B_\nu(T_\mathrm{e}$) the black body function. The sizes derived for our
compact sources range between 1 and 10~mpc. Thus, the centimeter sources we find
associated with class~II methanol masers have optical depths and emission
measures typical of \hchii\ regions. Higher resolution observations are needed
to determine if they also have typical \hchii\ region sizes.

In summary, we have identified a number of promising
candidates for \hchii\ regions associated with class~II methanol masers. 
Further observations at higher angular resolution are needed to study the
properties and fully characterize these regions.

\acknowledgments
\begin{small}
We thank the anonymous referee for his/her useful comments. We are very
grateful to the NRAO staff for their cheerful help and support during the
shared-risk observing period. SK acknowledges partial support from DGAPA project
IN101310, UNAM. This research made use of the NASA's Astrophysics Data System.
\end{small}


\begin{table*}[htb!]
\caption{3.6~cm and 1.3~cm continuum and 6.7~GHz methanol maser results}
\centering
\scriptsize
\begin{tabular}{l c c c c c c c c c c c c c}
\hline\hline\noalign{\smallskip}	

&\multicolumn{6}{c}{centimeter continuum\supa}
&
&\multicolumn{4}{c}{6.7~GHz methanol maser\supb}
\\
\cline{2-7}\cline{9-12}\noalign{\smallskip}

&$\alpha_\mathrm{(J2000.0)}$
&$\delta_\mathrm{(J2000.0)}$
&$S_{\nu; \mathrm{3.6~cm}}$\supc
&$S_{\nu; \mathrm{1.3~cm}}$\supc
&Spectral
&
&
&$\theta_\mathrm{offset}$\supf
&$\mathrm{offset}$\supf
&$S_\mathrm{peak}$
&$L_\mathrm{maser}$
\\
Region
&(~\raun~)
&(~\deun~)
&(mJy)
&(mJy)
&Index\supd
&Notes\supe
&
&(arcsec)
&(pc)
&(Jy)
&(\lo)
\\
\hline\hline
\noalign{\smallskip}
G34.82$+$0.35	&18 53 37.96		&01 50 30.4
		&$\phn0.18\pm0.03$	&$<0.35$		&$<0.87$
		&compact
		&&\phnn1.1	&0.02	&\phn0.24		&$8.98\times10^{-9}$	\\
		&18 53 38.69		&01 50 13.6
		&$\phn0.65\pm0.07$	&$\phn1.15\pm0.41$	&$+0.58\pm0.38$
		&compact
		&&\phn21.0	&0.37	&\phn0.24		&$8.98\times10^{-9}$	\\
G36.70$+$0.09	&\ldots			&\ldots
		&$<0.13$		&$<0.38$		&
		&
		&&		&	&\phn7.00		&$5.54\times10^{-6}$	\\
G37.55$+$0.19	&18 59 10.09		&04 12 17.5
		&$\phn0.60\pm0.08$	&$\phn0.67\pm0.13$	&$+0.11\pm0.24$
		&compact
		&&\phnn2.8	&0.07	&\phn5.27		&$1.40\times10^{-6}$	\\
		&18 59 10.34		&04 12 09.9
		&$<0.14$		&$\phn0.68\pm0.17$	&$>1.32$
		&compact
		&&\phnn7.4	&0.20	&\phn5.27		&$1.40\times10^{-6}$	\\
G37.60$+$0.42	&\ldots			&\ldots
		&$<0.14$		&$<0.32$		&
		&
		&&		&	&17.30			&$6.74\times10^{-6}$	\\
G38.03$-$0.30	&\ldots			&\ldots
		&$<0.30$		&$<0.36$		&
		&
		&&		&	&11.71			&$1.91\times10^{-6}$	\\
G38.08$-$0.27	&\ldots			&\ldots
		&$<0.36$		&$<0.35$		&
		&
		&&		&	&\phn0.59		&$2.10\times10^{-6}$	\\
G38.12$-$0.24	&19 01 43.73		&04 30 51.1
		&$22.42\pm2.24$		&$17.19\pm3.52$		&$-0.27\pm0.23$
		&large
		&&\phn15.1	&0.40	&\phn1.92		&$1.19\times10^{-6}$	\\
G38.26$-$0.08	&\ldots			&\ldots
		&$<0.20$		&$<0.39$		&
		&
		&&		&	&\phn7.03		&$5.56\times10^{-6}$	\\
G38.26$-$0.20	&19 01 54.4\phn		&04 39 26.\phn
		&$27.73$		&$<0.48$		&\ldots\supg
		&extended
		&&\phn51.3	&2.24	&\phn0.72		&$7.19\times10^{-7}$	\\
G38.92$-$0.36	&19 03 40.10		&05 10 14.4
		&$\phn5.65\pm0.57$	&$\phn4.25\pm0.94$	&$-0.29\pm0.25$
		&large
		&&\phn38.5	&1.96	&\phn1.26		&$9.86\times10^{-7}$	\\
G40.28$-$0.22	&19 05 41.20		&06 26 13.0
		&$\phn0.72\pm0.08$	&$\phn0.95\pm0.15$	&$+0.28\pm0.20$
		&compact
		&&\phnn0.4	&0.01	&24.47			&$1.13\times10^{-5}$	\\
G41.12$-$0.11	&\ldots			&\ldots
		&$<0.63$		&$<0.34$		&
		&
		&&		&	&\phn1.14		&$4.37\times10^{-7}$	\\
G41.16$-$0.20	&\ldots			&\ldots
		&$<1.41$		&$<0.34$		&
		&
		&&		&	&\phn0.27		&$1.05\times10^{-7}$	\\
G41.27$+$0.37	&\ldots			&\ldots
		&$<0.73$		&$<0.27$		&
		&
		&&		&	&\phn0.26		&$1.47\times10^{-7}$	\\
G42.03$+$0.19	&\ldots			&\ldots
		&			&$<0.33$		&
		&
		&&		&	&26.32			&$2.57\times10^{-5}$	\\
G42.30$-$0.30	&\ldots			&\ldots
		&			&$<0.33$		&
		&
		&&		&	&\phn6.33		&$3.87\times10^{-6}$	\\
G42.70$-$0.15	&19 09 55.16		&08 36 53.5
		&$\phn1.43\pm0.08$	&$\phn1.05\pm0.29$	&$-0.32\pm0.29$
		&compact
		&&\phnn1.4	&0.10	&\phn3.25		&$7.06\times10^{-6}$	\\
G45.81$-$0.36	&\ldots			&\ldots
		&$<0.38$		&$<0.34$		&
		&
		&&		&	&11.31			&$2.86\times10^{-6}$	\\
G48.99$-$0.30	&19 22 26.06		&14 06 39.1
		&$\phn7.63\pm0.69$	&$10.57\pm2.78$		&$+0.33\pm0.28$
		&compact
		&&\phnn1.3	&0.03	&\phn0.58		&$9.91\times10^{-8}$	\\
		&19 22 25.5\phn		&14 06 17.\phn
		&$1206.76$		&$352.4\pm0.21$		&\ldots\supg
		&extended
		&&\phn24.6	&0.64	&\phn0.58		&$9.91\times10^{-8}$	\\
G49.41$+$0.33	&19 20 59.83		&14 46 49.9
		&			&$\phn1.05\pm0.26$	&
		&large
		&&\phnn9.0	&0.53	&\phn9.25		&$2.39\times10^{-5}$	\\
G50.78$+$0.15	&\ldots			&\ldots
		&			&$<0.36$		&
		&
		&&		&	&\phn5.26		&$9.47\times10^{-7}$	\\
G52.92$+$0.41	&\ldots			&\ldots
		&			&$<0.43$		&
		&
		&&		&	&\phn6.64		&$8.44\times10^{-7}$	\\
G53.04$+$0.11	&19 28 55.53		&17 52 04.7
		&$\phn1.78\pm0.11$	&$\phn2.54\pm0.67$	&$+0.36\pm0.28$
		&compact
		&&\phnn1.7	&0.08	&\phn1.66		&$4.23\times10^{-7}$	\\
G53.14$+$0.07	&19 29 17.68		&17 56 19.2
		&$<0.24$		&$\phn1.78\pm0.54$	&$>1.68$
		&compact
		&&\phnn4.3	&0.04	&\phn1.02		&$1.35\times10^{-8}$	\\
\hline
\end{tabular}
\begin{list}{}{}
\item[\supa] Results for the 3.6~cm and 1.3~cm continuum emission located within
the 1.3~cm primary beam  ($\sim\!2$\arcmin). The pointing centers of the
observations correspond (in most cases) to the methanol maser positions given by 
Pandian \et\ (2011). Regions G42.03$+$0.19, G42.30$-$0.30,
G49.41$+$0.33, G50.78$+$0.15, and G52.92$+$0.11 were not observed at 3.6~cm 
because of missing scans during the shared-risk observations.
\item[\supb] Properties of the class~II methanol masers. Peak flux density and
luminosity of the maser components from Pandian \et\ (2007, 2009).
\item[\supc] Primary beam corrected. Uncertainties in flux density were calculated
assuming an uncertainty of 5\% and 25\% in the flux scale
determination at 3.6~cm and 1.3~cm, respectively. In case of non-detection, an
upper limit of 4 times the rms noise of the map is listed.
\item[\supd] Spectral index ($\alpha$: $S_\nu\propto\nu^\alpha$) estimated from
the flux densities at 3.6~cm and 1.3~cm, obtained with the \emph{uv}--range
common to both wavelengths, 3--28~k$\lambda$.
\item[\supe] Notes on the morphology of the centimeter emission, indicating if
it is compact (or unresolved), double, large (less compact and with some
structure) or extended.
\item[\supf] Angular (in arcsec) and spatial (in pc) offsets between the radio
continuum source with the coordinates indicated in this Table, and the position
of the methanol maser (Pandian \et\ 2011). The distance used to determine the
spatial offset is reported in Pandian \et\ (2009).
\item[\supg] The source is filtered out at 1.3~cm, resulting in an unreliable
spectral index determination.
\end{list}
\label{t:cmmaser}
\end{table*}

\end{document}